\documentclass{aa}
\def\b{\vec}
\def\ueber#1#2{{\setbox0=\hbox{$#1$}%
  \setbox1=\hbox to\wd0{\hss$\scriptscriptstyle #2$\hss}%
  \offinterlineskip
  \vbox{\box1\kern0.4mm\box0}}{}}
\def\etal{{et al.~}}
\begin{document}
\thesaurus{02(12.03.4; 12.12.1)}
\title{Cosmic density and velocity fields\\
in Lagrangian perturbation theory}

\author{Mikel Susperregi\inst1 \and
Thomas Buchert\inst2}

\institute{Astronomy Unit, School of Mathematical Sciences,
Queen Mary
\& Westfield, London E1 4NS, UK (mps@maths.qmw.ac.uk)\and
Theoretische Physik, Ludwig-Maximilians-Universit\"at,
Theresienstr. 37,
D-80333 M\"unchen, Germany\\
(buchert@stat.physik.uni-muenchen.de)}
\date{Received 23 May 1996 / Accepted 16 December 1996}
\authorrunning{M. Susperregi \& T. Buchert}
\maketitle

\begin{abstract}
A first- and second-order relation between cosmic density
and peculiar-velocity fields is
presented. The calculation is purely Lagrangian and it is derived using the
second-order solutions of the Lagrange-Newton system obtained by
Buchert \& Ehlers. The procedure is applied to two particular solutions
given generic initial conditions. In this approach, the continuity
equation yields a relation between the over-density
and peculiar-velocity fields that automatically satisfies Euler's equation
because the orbits are derived from the Lagrange-Newton system.
This scheme generalizes some results obtained by Nusser et al. (1991) in
the context of the Zel'dovich approximation. As opposed to several other
reconstruction schemes, in this approach it is not necessary to truncate
the expansion of the Jacobian given by the continuity equation
in order to calculate a first- or second-order expression for the density field.
In these previous schemes, the density contrast given by (a) the continuity
equation and (b) Euler's equation are mutually incompatible.
This inconsistency arises as a consequence of an improper handling of
Lagrangian and Eulerian coordinates in the analysis.
Here, we take into account the fact that an exact
calculation of the density is
feasible in the Lagrangian picture
and therefore an accurate and consistent description is
obtained.
\keywords{cosmology: theory --
large-scale structure of the universe}
\end{abstract}
\section{Introduction}
An important aim of cosmology is to determine a relationship
between the cosmic density contrast $\delta$ and the peculiar-velocity
field $\b u$ under the assumption that these fields have evolved
under the action of gravity.
In the Eulerian theory of gravitational instability,
retaining only the linear growing-mode solution,
this relationship is simply given by\footnote{Throughout
this article, $\b q$ and $\b X$ denote (comoving)
Eulerian and Lagrangian coordinates
respectively.}
\begin{equation}\label{linear}
\delta(t,\b q)=- {D\over a\dot D}\;{\nabla_{q}}\cdot\b u(t,\b q),
\end{equation}
where $D(t)$ is the
usual linear growth factor, such that $D(t)=a(t)=(t/t_0)^{2/3}$ in an
Einstein-de Sitter universe. Beyond linear theory, in the weakly non-linear
regime, (\ref{linear}) has been used to good effect as an
estimate of the density contrast field from observed peculiar-velocities (e.g.
Bertschinger \& Dekel 1989; Dekel et al. 1990;
Bertschinger et al. 1990). On the other hand, the POTENT method
and other quasi-linear reconstruction procedures involve the
Zel'dovich approximation (ZA)(Nusser et al. 1991; Nusser \& Dekel 1992).
The ZA does not however, strictly speaking,
provide an exact algebraic relation between the fields
$\b u (t,\b q)$ and $\delta (t,\b q)$. It can however yield (as shown in
Nusser \etal 1991) a relation between the fields that
is self-consistent within
the order of the approximation, i.e. within $O(D^2)$.
In the ZA, from the Lagrangian integral of the continuity equation we have

\begin{equation} \label{ZArecons}
{1+\delta(t,\b X)\over 1+\delta(t_0,\b X)}=\det\biggl[{\delta_{ij}}
+{D\over a \dot D}
{\partial u_i (t,\b X)\over\partial X_j}\biggr]^{-1}\;,\label{zeldovich}
\end{equation}

\noindent
and the fields are mapped back to Eulerian
space via the ZA map for the orbits of
the particles. Initial conditions are specified at an arbitrary
early time $t_0\geq 0$. Also, it is customary to make the assumption of
smoothness at early times, $\delta(t_0,\b X)\approx 0$, by virtue of
the amplitude of the fluctuations in the microwave background.
From (\ref{ZArecons}) we obtain the linear relation (\ref{linear})
by truncating the expansion of the determinant (\ref{zeldovich}) to the
lowest order in the perturbation field $\b u$
and replacing Lagrangian by Eulerian coordinates ($\b X \approx \b q$).

\noindent
These schemes have been generalized further (Gramann 1993a, 1993b; for
their Eulerian analogue, see Bernardeau 1992) by starting out with a
parametrization of the particle orbits $\b q=\b F(t,\b X)$
of the type:

\begin{equation}
\b F(t,\b X)= \b X+ D\nabla_{X}\Phi^{(1)}(\b X)+D^2\nabla_{X}
\Phi^{(2)}(\b X), \label{moutarde}
\end{equation}

\noindent In the case of a flat Universe, the leading
second-order term in perturbation theory (Bouchet et al.
1992) is indeed $\propto D^2(t)$.
Therefore, the coordinate map (\ref{moutarde}) is
in principle well-motivated (for a numerical comparison of the
different models, see e.g. Dekel 1994). However, as shown in
Gramann (1993b), orbits of the type (\ref{moutarde}) do not yield
a self-consistent relation for the cosmic density and velocity
fields; the density fields obtained via (a) the continuity equation and
(b) the Euler equation differ within the order of the approximation.
The reason for this is that, as we shall see, (\ref{moutarde})
is only an approximate second-order solution
of the Lagrange-Newton system.
Thus, we wish to call the attention of the
reader to the fact that the reconstruction models
derived from (\ref{moutarde}) and, more
generally, from an ansatz including higher orders in $D(t)$,

\begin{equation}   \label{D}
\b F(t,\b X)= \b X+ \sum_{n=1}^N D^n\,\nabla_X\Phi^{(n)}(\b X)
\end{equation}

\noindent (for arbitrary N),
do not follow a rigorous line of analysis (in the
case of an arbitrary $\Omega$) for the following reasons:

\newcounter{bean}
\begin{list}{\Roman{bean}}{\usecounter{bean}}
 \item In these models, the parametrization of the orbits is given {  ad hoc}
and it is not derived as a solution of the Lagrangian evolution
equations for the flow field. In particular,
the absence of lower-order growing
modes, that are present in the general perturbation theory
solution, implies that the derivation of $\delta$ through
the continuity equation neglects couplings
of these modes with leading-order terms.
 \item The density contrast $\delta_c$ obtained from the continuity equation
in perturbation theory and $\delta_e$ satisfying Euler's equation are
mutually incompatible.
 \item There is no reason to justify that $D(t)$
is a good perturbative parameter (in terms of the convergence of the solutions)
and that a meaningful generalization of
the family of solutions where the ZA
belongs can be realized as a polynomial in $D(t)$.
  \end{list}

\noindent It is easy to see that (II) follows as a consequence of (I).
Regarding (I) and (III), it has been shown that the correct
coordinate map between Eulerian and Lagrangian coordinates in
Lagrangian perturbation theory
is obtained by calculating the perturbative solution of
the Lagrange-Newton system for the trajectories $\b q=\b F(t,\b X)$.
These evolution equations are obtained by transforming
the Euler-Newton system to Lagrangian coordinates and eliminating
all Eulerian fields by using exact Lagrangian integrals for the Eulerian
acceleration and density (Buchert \& G\"otz 1987; Buchert 1989). Within
this scheme, the ZA is recovered as a
subclass of solutions (Buchert 1992), and hence, one
can obtain a self-consistent solution for the over-density
field in terms of the velocities as in Nusser \etal (1991).
Furthermore, the second- and third-order solutions
obtained in the Lagrangian framework
(Buchert \& Ehlers 1993, Buchert 1994; and in a slightly different approach
Bouchet et al. 1992, 1995; Catelan 1995)
are unique and well-defined at all times from a fiducial time $t_0$,
where initial data are specified,\footnote{We make use of the term
`initial' to denote conditions given at $t_0$, an arbitrary but
sufficiently early epoch, not necessarily $t=0$.}
until shell-crossing, provided we impose periodic boundary conditions and
fix some global gauge conditions
(Ehlers \&
Buchert 1997)
as opposed to the orbits described by polynomials
in $D(t)$, which are expected to
have poor convergence (for a discussion in the context of the
variational approach to the solutions see Susperregi \& Binney 1994).
Although the leading terms of the longitudinal parts of the
perturbation solutions seem to confirm this polynomial approach
(e.g. Bouchet et al. 1992, for the second-order solutions),
this is not a mathematically consistent motivation for
 using $D(t)$ as an expansion parameter
by going to higher orders in perturbation theory. On the contrary,
the structure of the Lagrangian perturbation
scheme is such that, starting at third order, the presence of
interaction terms among perturbations of different orders are an obstacle
to convergence,
and in fact for $n>3$ all terms are interaction terms (see Buchert 1994).
Therefore, for $n\geq 3$ an ansatz of the form (\ref{D}) cannot
reproduce gravitational dynamics, as stated in (III).
This is an especially
critical point to bear in mind in the context of reconstruction models
in which higher-order couplings of flow field gradients show up
in the density-velocity relation.

\noindent
In this paper, we approach the reconstruction problem from a purely
Lagrangian perspective using the solutions of the Lagrange-Newton
system obtained by Buchert \& Ehlers (1993). The goal of the
paper is to obtain a self-consistent solution for the over-density
field that is valid to second order, thus extending the results
of Nusser \etal 1991 in the context of the ZA. Our results are
purely Lagrangian, but in obtaining them we rely on the fact that
there is a one-to-one mapping between Lagrangian and Eulerian
coordinates and hence, the reconstruction formulae presented are
only valid up to orbit-crossing time. The basic
formalism is laid out in Sect. 2,
and in Sect. 3 we discuss its application to two different classes
of initial conditions. In Sect. 4 we discuss the self-consistency
of solutions in perturbation theory models.
Sect. 5 compares the model presented with
previous reconstruction procedures
and Sect. 6 sums up.

\section{First- and second-order Lagrangian field reconstruction}

In what follows, we shall mainly discuss the physically interesting
case where the peculiar-velocity field is parallel to the
peculiar-acceleration, up to decaying modes, a property which results from
the Eulerian linear theory for irrotational fields, and is
also valid for the general irrotational Lagrangian linear solution
(the growing-mode
solution supports the tendency to becoming parallel, see Buchert (1992)).
We will further recall that
both fields are practically parallel up to orbit-crossing time,
also in the second-order schemes we will consider. The results
presented however are not restricted to solutions where this property
holds but, as we will see below, apply to a wider class of solutions.

In what follows we shall study subclasses of the second-order
Lagrangian perturbation solutions given by Buchert \& Ehlers (1993)
which are only restricted by the requirement that there exists
an arbitrary functional relationship among peculiar-velocity and
-acceleration at the initial time. The subclasses are singled out
by special assumptions for this functional relationship.

The second-order parametrization of the orbits of the particles
in terms of their Lagrangian
coordinates and time for the subclasses of solutions
which we consider later can be written as follows:

\begin{equation}
\b q=\b F(t,\b X)= \b X +f_1(t) \nabla_{X} S^{(1)}(\b X)+f_2(t)
\nabla_{X} S^{(2)}(\b X),
\label{eulerian}
\end{equation}

\noindent where $f_1(t)$ and $f_2(t)$ are dimensionless functions of time,
calculated from solving the Lagrange-Newton system
for deviations from the homogeneous solution to second order.
Moutarde et al. (1991) investigated the orbits (\ref{eulerian})
(rather than (\ref{D})),
for leading orders of $f_1(t)$ and $f_2(t)$ and for a model of
a particular initial potential.
In the next section, we will examine two particular solutions for
$f_1(t)$ and $f_2(t)$ for different, but generic, initial conditions.

The expansion factor $a(t)$ is normalized to unity at $t_0$,
$a(t)=\Bigl({t\over t_0}\Bigr)^{2/3}$,
(and equal to the linear growth factor $D(t)$ in
an Einstein-de Sitter universe considered here).
The perturbation potentials $S^{(1)}$ and $S^{(2)}$
are determined by the initial data such that, in general, they have to
satisfy the (non-local) boundary conditions:

\begin{equation}
\nabla_{X}^2 S^{(1)}=I(S)\,t_0, \label{s1}
\end{equation}

\begin{equation}
\nabla_{X}^2 S^{(2)}=2\,II(S^{(1)}), \label{s2}
\end{equation}

\noindent 
where $S$ denotes the initial peculiar-velocity potential

\noindent
$\b u(t_0,\b X)$ $\equiv \nabla_{X} S(\b X)$,
and (following the summation convention)

\begin{equation}
I(S)=S_{,i,i},
\end{equation}

\begin{equation}
II(S^{(1)})\equiv {1\over 2}\;
(S^{(1)}_{,i,i}S^{(1)}_{,j,j}-S^{(1)}_{,i,j}S^{(1)}_{,j,i}),
\end{equation}

\noindent are the first and second
principal invariants of the peculiar-velocity tensor gradient.
For periodic boundary conditions and suitable gauge conditions, however,
which we shall assume hereafter, we can
simplify (\ref{s1})(\ref{s2})
(see Ehlers \& Buchert (1997) for a rigourous proof):

\begin{equation}
\nabla_{X} S^{(1)}=\,\nabla S \,t_0\;, \label{pbc1}
\end{equation}

\begin{equation}
\nabla_{X}^2 S^{(2)}=2\,II(S)\,t_0^2\;.  \label{pbc2}
\end{equation}

\noindent Therefore, (\ref{eulerian}) gives us the trajectory of
any given particle by
solving (\ref{pbc1})(\ref{pbc2}) for an initial potential $S$.
At the time $t_0$,
both $f_1$ and $f_2$ vanish,
so Lagrangian and Eulerian coordinates coincide,

\begin{equation}
\b q = \b F(t_0, \b X)=\b X.
\end{equation}

\noindent Thus, the expression for the density contrast
at any given time and for any trajectory field reads:

\begin{equation}
[\,1+\delta(t,\b X)\,]=
[\,1+\delta(t_0,\b X)\,]\;J^{-1}\;, \label{density1}
\end{equation}

\noindent where the determinant $J(t,\b X)$ is

\begin{equation}
J\equiv \det \Biggl({\partial F_i\over\partial X_j}\Biggr) \label{det1}.
\end{equation}

\noindent
In order to calculate $\delta$ as a function of Eulerian coordinates,
we need to invert the coordinate map (\ref{eulerian}) so that
Lagrangian coordinates
are expressed in terms of the Eulerian coordinates and thereby
derive the following:

\begin{equation}
[\,1+\delta(t,\b H (t,\b q))\,]=[\,1+\ueber{\delta}{o}\,]\;J^E ,
\label{density2}
\end{equation}

\noindent where $\ueber{\delta}{o}(\b X)$ is the initial density contrast
and the determinant $J^E (t,\b q)= J^{-1} (t,\b X)$ is

\begin{equation}
J^E\equiv \det \Biggl({\partial H_i\over\partial q_j}\Biggr) \label{det2}.
\end{equation}

\noindent Eq. (\ref{density2}) gives the Eulerian values of the
field $\delta$ at the Eulerian positions corresponding to a given trajectory
labelled by $\b X=\b H(t,\b q)$.
In principle, we can interpret this as a field
$\delta (t,\b q)$ satisfying the Eulerian equations of motion.
Nonetheless, it is important to bear in mind that
the functional dependence of $\delta$ on $\b H$ is only well-defined
while the map between both coordinate spaces remains one-to-one.
The peculiar-velocity is given by

\begin{equation}
\b u(t,\b H) = a\dot a\; {{\mathrm{d}}\over{\mathrm{d}} a}\b F\;\;,   \label{u}
\end{equation}

\noindent and, similarly, the peculiar-acceleration,

\begin{equation}
\b w(t,\b H)
= (a\ddot a+2\dot a^2){{\mathrm{d}}\over{\mathrm{d}} a}\b F +
a\dot a^2{{\mathrm{d}}^2\over{\mathrm{d}} a^2} \b F \;\;. \label{w}
\end{equation}

\noindent The Jacobian $J^E$ of the transformation $\b H$ is given
explicitly in Appendix B.
This general expression involves a number of cross-terms
of $\b u$ and $\b w$ components. However, we can ignore these by assuming
the peculiar-velocity field to be parallel to the
peculiar-acceleration, such that

\begin{equation}
\b w(t,\b H)= \b u(t,\b H)\,t^{-1}. \label{parallelism}
\end{equation}

\noindent This assumption can be justified on physical grounds:
after some time, the peculiar-velocity
field tends to be parallel to the peculiar-acceleration,
for the second-order irrotational solution (\ref{eulerian}),
due to the existence of decaying and growing solutions in the
weakly non-linear regime. We will address this issue in detail
in Sect. 5, and show that a weaker condition on the
gradients of the fields leads to the same result.
As in the first-order model (where (\ref{parallelism})
holds exactly) the growing part supports
parallelism, as it is shown in Buchert \& Ehlers (1993) for a wide
class of irrotational second-order
solutions. Indeed, the difference
between $\b u \,t^{-1}$ and $\b w$
for (\ref{eulerian}) is given by decaying modes
only, as we shall see in the next section.
Therefore, we are entitled to express $\b w$ by some function
proportional to $\b u$,
so that the inversion of the map (\ref{eulerian}) satisfies
the implicit equation:

\begin{equation}
\b H(t,\b q)=\b q-h(t)\;\b u(t,\b H(t,\b q)),  \label{lagrangian2}
\end{equation}

\noindent where $h(t)$ is given explicitly in Appendix A for
two different models.

In the first-order model, by virtue of
(\ref{parallelism}), the same relation (\ref{lagrangian2}) holds
with the appropriate $h(t)$ (Appendix A.1). With condition (\ref{parallelism})
implicit in (\ref{lagrangian2}), the determinant $J^{E}$
reduces to:

\begin{equation}
J^E=1-h\;I^E(u_i)+h^2\;II^E(u_i)-h^3\;III^E(u_i). \label{jacobian}
\end{equation}

\noindent Here, $I^E$ (divergence), $II^E$ (dispersion of diagonal
compo\-nents)
and $III^E$ (determinant)
denote the three principal scalar invariants of
the peculiar-velocity gradient with respect to comoving
Eulerian coordinates as given in Appendix B.
Therefore, substituting (\ref{jacobian}) into (\ref{density2}),
we derive the relation for the density and velocity fields:
\[
{1+\delta(t,\b H (t,\b q))\over 1+\ueber{\delta}{o}}=
\]
\begin{equation} \label{result}
\hspace{0.5cm}1-h\;I^E(u_i)+h^2\;II^E(u_i)-h^3\;III^E(u_i)\;,
\end{equation}

\noindent where $u_i = u_i (t,\b H (t,\b q))$. This
result is unique and valid up to orbit-crossing time, i.e.,
while the coordinate inversion $\b X = \b H(t,\b q)$
is well-defined. In principle, it enables us to calculate the cosmic density
field from observations of the peculiar-velocity field
in the weakly non-linear regime at a fixed time, provided
we set $\delta(t_0)=0$.
It is interesting to emphasize that,
due to the parallelism condition (\ref{parallelism}),
(\ref{result}) applies at both first- and second-order (the appropriate
$h(t)$ for each case is given in Appendix A).
The simplicity of (\ref{result}) is remarkable.
It shows that the first- and second-order models
contain not only quadratic terms
in $u_{i,j}$ (as pointed out in the literature, e.g. Gramann 1993a,b)
but also cubic terms, arranged in the form of the
principal invariants of
the deformation tensor: $I^E(u_i)$, $II^E(u_i)$ and $III^E(u_i)$.

A more general result, albeit a complicated one, can also be derived
for the case where we drop the condition (\ref{parallelism}).
The corresponding solution, taking into account all cross-terms,
is obtained using the full solution for $J^E$ given in Appendix B.
Nonetheless, considering $\b w$ is not directly measurable from observations,
the application of the general solution without the assumption of
parallelism would be impractical. In these circumstances
(\ref{result}) provides a useful approximation.

\section{Application to specific models}

In this section, we will examine two particular solutions of
the second-order orbits
(\ref{eulerian}) that are of interest in the reconstruction
procedure described in Sect. 2. These are characterized by the
nature of the initial conditions at $t_0$. At this time, only one
field, S, is specified. These solutions are derived in Buchert \&
Ehlers (1993).

\vspace{0.6cm}\noindent
{\it 3.1. Model I: $\vec{u}(\vec{X})=\b w(\b X)t_0$}\hfill\break
\vspace{0.3cm}

\noindent
In this case, the velocity and acceleration fields are set to
be identically parallel at the initial time. This relation
is satisfied in Zel'dovich's ansatz, and it is a natural
condition to set in a more general scheme, given the
remarkable accuracy of the ``Zel'dovich approximation''. The initial density
perturbation is thus expected to be proportional to the
divergence of the peculiar-velocity.
The solution (\ref{eulerian}) then is specified by the functions

\begin{equation}
f_1(t)={3\over 2} [\,a(t)-1\,],
\end{equation}

\begin{equation}
f_2(t)={9\over 4}[-{3\over 14}a^2(t)+{3\over 5}a(t)-
{1\over 2}+{4\over 35}a^{-3/2}(t)\,].
\end{equation}

\noindent Therefore, the peculiar-velocity reads
\[
\b u(t,\b H)= a^{1/2}\,t_0^{-1}\,\nabla_X S^{(1)}
\]
\begin{equation}
\hspace{0.5cm} +{3\over 2}(-{3\over 7}a^{3/2}+{3\over 5}a^{1/2}-{6\over
35}a^{-2})
\,t_0^{-1}\,\nabla_X S^{(2)},
\end{equation}

\noindent and the peculiar-acceleration,
\[
\b w(t,\b H)= a^{-1}\,t_0^{-2}\,\nabla_X S^{(1)}
\]
\begin{equation}
\hspace{0.5cm} +(-{15\over 14} +{9\over 10}a^{-1}+{6\over
35}a^{-7/2})\,t_0^{-2}\,
\nabla_X S^{(2)}.
\end{equation}

\noindent Thus the departure from the parallelism condition (\ref{parallelism})
is

\begin{equation} \label{diff-I}
\b u\, t^{-1}-\b w
= {3\over 7}\,(1-a^{-7/2})\,t_0^{-2}\;\nabla_{X} S^{(2)}.
\end{equation}

\noindent Hence, this difference is a negligible effect
due to the presence of the decaying mode. At $t=t_0$,
the RHS of (\ref{diff-I}) is identically zero and, at
later times, the decaying mode vanishes sufficiently fast
and the overall coefficient therefore tends to a constant.
This constant second-order term could be absorbed
in a coordinate transformation. In this case, $h(t)$ in
the reconstruction formula (\ref{result}) is given by
(see Appendix A.2 for details)

\begin{equation}
h= {3\over 2}\,t_0\;(1-a^{-7/2})^{-1}(a^{1/2}-a^{-1/2}-a^{-3}+a^{-4}).
\end{equation}
This is a monotonic function of $t$, and to the leading order it
reads

\begin{equation}
h\approx {3\over 2}\,t_0\; a^{1/2},
\end{equation}

\noindent which is consistent with the results presented in
Moutarde et al. (1991).

\vspace{0.6cm}\noindent
{\it 3.2. Model II: $\vec{w}(\vec{X})=0$}
\vspace{0.3cm}

\noindent
Although Model I provides an appropriate initial value setting
from the physical point of view, the initial density contrast
in this model is strictly non-vanishing. However, for reconstruction
models we have to neglect this, since we do not want to specify
initial information. Precisely speaking, this implies an error
which can be removed by using Model II:
Given a vanishing initial peculiar-acceleration field,
the initial density contrast $\ueber{\delta}{o}$
is zero and its growth is
induced by the initial velocity perturbations only. In this case,
$f_1(t)$ and $f_2(t)$ in the class of solutions for the orbits (\ref{eulerian})
are given by

\begin{equation}
f_1(t)={3\over 5}\,[\,a(t)-a^{-3/2}(t)\,],
\end{equation}
\[
f_2(t)={9\over 25}\,[\,-{3\over 14}a^2(t)+{29\over 40}a(t)
-{1\over 2} -{1\over 2} a^{-1/2}(t) \]
\begin{equation}
\hspace{0.5cm}+{43\over 70}a^{-3/2}(t)-{1\over 8} a^{-3}(t)\,].
\end{equation}

\noindent Therefore, from (\ref{u}) the peculiar-velocity is

\[
\b u(t,\b H)= {2\over 5}(a^{1/2}+{3\over 2} a^{-2})\,t_0^{-1}\,
\nabla_{X} S^{(1)} +{9\over 25}\,(-{2\over 7}a^{3/2}
\]
\[
+{29\over 60}a^{1/2}+{1\over 6}a^{-1}
 -{43\over 70}a^{-2}+{1\over 4} a^{-7/2})\,t_0^{-1}\,\nabla_{X} S^{(2)},
\]
\begin{equation}
\end{equation}

\noindent and the peculiar-acceleration,

\[
  \b w(t,\b H)={2\over 5} \, (a^{-1}-a^{-7/2})\,t_0^{-2}\,\nabla_{X} S^{(1)}
\]
\begin{equation}
+{3\over 25}\,(-{10\over 7}+{29\over 20}a^{-1}+{43\over 35}a^{-7/2}
 -{5\over 4}a^{-5})\,t_0^{-2}\,\nabla_{X} S^{(2)}.
\end{equation}

\noindent Like in the previous example we can calculate
the departure from parallelism which is given in terms of the decaying
modes
\[
 \b u t^{-1}-\b w = a^{-7/2}\,t_0^{-2}\,\nabla_{X} S^{(1)}
\]
\begin{equation} \label{diff-II}
\hspace{0.5cm}+{3\over 25}({4\over 7}-{1\over 2}a^{-5/2}-
{43\over 14} a^{-7/2}+2 a^{-5})\,t_0^{-2}\,\nabla_{X} S^{(2)},
\end{equation}

\noindent such that, it is once more manifest that
at late times parallelism is supported by the
growing modes. At first sight it might seem that the presence
of the first-order potential in (\ref{diff-II}) is a shortcoming
for model II in comparison to model I, where departure from parallelism
is a second-order effect, given by (\ref{diff-I}).
However, although the initial
fields are non-parallel as indicated by (\ref{diff-II}),
the orders of magnitude involved in the RHS are very small
from the outset and
these decaying modes soon decrease further to negligible values.

\vspace{0.6cm}\noindent
{\it 3.3. Parallelism of ${\b u}$ and ${\b w}$}
\vspace{0.3cm}

\noindent
The assumption of parallelism (\ref{parallelism}) is justified by
the presence of decaying modes in the difference $t\,\b w-\b u$ in
both models I and II. On the other hand, it is apparent that the
exact expression for $J^E$ requires knowledge of the derivatives of
the peculiar-acceleration field. Using the parallelism condition
to simplify the solution implies neglecting the contribution of the
Eulerian derivatives of the vector field $t\, \b w-\b u$. In this
section we produce an analytical argument to justify this
approximation in the weakly nonlinear regime.

A measure of parallelism of ${\b u}$ and ${\b w}$ is given by
the quantity $|\b u\times\b w|$ (an alternative measure has been proposed
recently by Bagla \& Padmanabhan (1996)).
We expect this quantity to be
small in the regime where $\b u$ and $\b w$ are close to being
parallel. As is given explicitly in Appendix A, we have

\begin{equation}
\b u(t,\b H)=\alpha(t)\nabla_{\b X}S^{(1)}+\beta(t)
\nabla_{\b X}S^{(2)},
\end{equation}

\begin{equation}
\b w(t,\b H)=\zeta(t)\nabla_{\b X}S^{(1)}+\eta(t)\nabla_{\b X}S^{(2)},
\end{equation}

\noindent and therefore,

\begin{equation}
\b u\times\b w=\big(\alpha\,\zeta-\beta\,\eta\big)\;
\nabla_{\b X}S^{(1)}\times\nabla_{\b X}S^{(2)}.
\end{equation}

\noindent For the particular case of Model I, for instance, this leads to

\begin{equation}
\mid\b u\times\b w\mid={3\over 7}t_0^{-3}\;a^{1/2}\;\mid 1-a^{-7/2}\mid
|\nabla_{\b X}S^{(1)}\times\nabla_{\b X}S^{(2)}|.
\label{wedge}
\end{equation}

\noindent This quantity vanishes for all models where $\nabla_{\b X}S^{(1)}$
and $\nabla_{\b X}S^{(2)}$ are parallel.
%, e.g. the initial condition
%$S={1\over 2}aX^2$ is such a case.
However this is not so in general,
and the integrals of (\ref{s1})(\ref{s2}) can yield to
non-parallel (and non-negligible)
gradients $\nabla_{\b X}S^{(1)}$ and $\nabla_{\b X}S^{(2)}$.
Furthermore, the presence of the coefficient $a^{1/2}(t)$ in
(\ref{wedge}) indicates that an initial departure from parallelism
of the fields $\b u$ and $\b w$ will grow in time. At this
point it is important to stress that our simplification
to derive (\ref{result}) does not strictly speaking rely on the
parallelism of the fields $\b u$ and $\b w$ but on that of the
spatial gradients $u_{i,j}$ and $w_{i,j}$. The differences of these
fields are intimately connected, but the time dependence involved is
different as is clear by direct differentiation. Parallelism
of the gradients is a weaker condition to impose than that of the fields
themselves. It is easy to show that
\[
u_{i,j}=\biggl[\alpha\;\nabla_{ik}S^{(1)}
+\beta\;\nabla_{ik}S^{(2)}\biggr]
\]
\begin{equation}
\hspace{0.5cm}\biggl[\delta_{kj}+f_1\;\nabla_{kj}S^{(1)}
+f_2\;\nabla_{kj}S^{(2)}\biggr]^{-1},
\end{equation}

\noindent and
\[
w_{i,j}=\biggl[\zeta\;\nabla_{ik}S^{(1)}
+\eta\;\nabla_{ik}S^{(2)}\biggr]
\]
\begin{equation}
\hspace{0.5cm}\biggl[\delta_{kj}+f_1\;\nabla_{kj}S^{(1)}
+f_2\;\nabla_{kj}S^{(2)}\biggr]^{-1},
\end{equation}

\noindent Therefore, we have that for Model I:

\begin{equation}
u_{i,j}\,t^{-1}-w_{i,j}={3\over 7}\big(1-a^{-7/2}\big)\,\nabla_{ik}S^{(2)}
\;\Gamma_{kj},
\end{equation}

\noindent and for Model II:
\[
u_{i,j}\,t^{-1}-w_{i,j}={3\over 5}a^{-7/2}\,\nabla_{ik}S^{(1)}\;\Gamma_{kj}
\]
\begin{equation}
\hspace{0.5cm}+{3\over 25}\Big({4\over 7}+{1\over 2}a^{-5/2}-{43\over
14}a^{-7/2}
-{1\over 2}a^{-5}\Big)\,\nabla_{ik}S^{(2)}\;\Gamma_{kj},
\end{equation}

\noindent where

\begin{equation}
\Gamma_{ij}\equiv\big[\delta_{ij}+f_1\nabla_{ij}S^{(1)}
+f_2\nabla_{ij}S^{(2)}\big]^{-1}.
\end{equation}

\noindent In the case of Model I, parallelism is exact in the initial
conditions, therefore there is no dependence of the derivatives on the
potential $S^{(1)}$. In all other models, the difference
$u_{i,j}\,t^{-1}-w_{i,j}$ comes as a linear combination of the
double derivatives $\nabla_{ij}S^{(n)}$, $n=1,2$, up to the coefficient
$\Gamma_{ij}$ that plays a role when we approach orbit-crossing.
For realistic fields, i.e., those that evolve from negligible amplitudes
at $t=t_0$, the gradients of the perturbative potentials
$S^{(1)}$, $S^{(2)}$ are sufficiently smooth and slowly varying, and
therefore they are well under the order of the approximation
in the weakly nonlinear regime.
We stress, however, as indicated above that in the strongly nonlinear regime
(or at the epoch of shell-crossing, respectively), this argument breaks down
in the same manner as perturbative solutions at any order do.

\section{Self-consistency of perturbative solutions}

In our reconstruction procedure we have proposed a simplification
of the exact solution (\ref{density2})
under the assumption of parallelism
of the fields $\b u$ and $\b w$. Thus we are able to
obtain an expression for the density in terms of the derivatives of
the velocity only. We have seen that this assumption is sufficiently
accurate, the difference between both fields is a negligible
effect. However, it is not necessary to adopt parallelism,
and in this section we will outline a possible scheme
to construct iteratively a ``self-consistent''
solution for $\b w$ from the velocity field, so that the
reconstruction method with the Jacobian (\ref{full-jacobian})
can be employed.

At this point, we would like to call the attention of the reader to
the problem of self-consistency of approximations as pointed out
by Doroshkevich et al. (1973) (compare the discussion by
Buchert (1989)). We enunciate the problem by
stating that perturbative solutions of the equations of motion
are mutually consistent only within the truncation order. This
can be illustrated with the ``Zel'dovich approximation''.
The density contrast as
given by the Poisson equation is

\begin{equation}
\delta^{(1)}(t,\b q)= -{3\over 2}\,t_0^2\,a^{-1}\,
\nabla_q \cdot\b w(t,\b q),
\end{equation}

\noindent and the density contrast as given by mass conservation

\begin{equation}
1+\delta^{(2)}(t,\b H)={1+\ueber{\delta}{o}\over J(t,\b H)}.
\end{equation}

\noindent Introducing the dimensionless error

\begin{equation}
\epsilon={\delta^{(2)}-\delta^{(1)}\over \delta^{(2)}},
\end{equation}

\noindent this difference for the ``Zel'dovich approximation'' gives

\begin{equation}
\epsilon=(a-1)^2\;II(S)+(a-1)^3\;III(S),
\end{equation}

\noindent which is within the order of the approximation. In
the second-order model we expect an analogous situation. As the
orbits we have used are obtained by solving the Lagrange-Newton
system perturbatively, Eulerian quantities constructed from these
will automatically satisfy all equations to second-order. However,
the density contrast constructed from the velocities through
(\ref{result}) is associated to a Newtonian potential causing
an acceleration that will in general differ from the acceleration
given by the second-order orbits. This difference vanishes only for
exact solutions.

Let us denote by a (1) superscript the quantities derived from the
particle orbits. Hence,
\[
{1+\delta^{(1)}(t,H)\over 1+\ueber{\delta}{o}}=
\]
\begin{equation} \label{iteration1}
\hspace{0.5cm}-h(t)\,I^E(u_i^{(1)})+h^2(t)\,II^E(u_i^{(1)})-h^3\,III^E(u_i^{(1)}),
\end{equation}

\noindent and therefore,
\[
\nabla_q \cdot\b w^{(2)}= -{2\over 3 t_0^2} a
\,(1+\ueber{\delta}{o})
\]
\begin{equation}
\hspace{0.5cm}\cdot
\lbrack -h\,I^E(u_i^{(1)})+h^2\,II^E (u_i^{(1)})-h^3\,III^E
(u_i^{(1)})-1 \rbrack\;,
\end{equation}

\noindent which in general differs from
$\b w^{(1)}={\dot{\b u}}^{(1)}+{\dot a \over a}{\b u}^{(1)}$. From
$\b w^{(2)}={\dot{\b u}}^{(2)}+{\dot a \over a}{\b u}^{(2)}$
it is possible in principle to solve for $\b u^{(2)}$,
from which $\delta^{(2)}$ is similarly constructed, and we
iterate henceforth via the prescription

\[
{1+\delta^{(n)}(t,H)\over 1+\ueber{\delta}{o}}=
\]
\begin{equation}
\hspace{0.5cm}
-h(t)\,I^E(u_i^{(n)})+h^2(t)\,II^E(u_i^{(n)})-h^3\,III^E(u_i^{(n)}),
\end{equation}

\noindent that we use to integrate $\b w^{(n+1)}$. A consistent
solution is then found in the limit

\begin{equation}
\epsilon = {\delta^{(n+1)}-\delta^{(n)}\over \delta^{(n)}}
\rightarrow 0\quad\quad\quad n\rightarrow\infty.
\end{equation}

\noindent Note that after each iteration, the condition
$\nabla_q \times\b w^{(n)}= {\b 0}$ (arising from the Lagrange equations
in the exact case) is preserved at all times. Finding a
self-consistent
solution to the dynamics
by successive iterations is tantamount to minimization of the
difference of the particle orbits

\begin{equation} \label{deltaF}
\Delta\b F(t,\b X)=\b F_{\mathrm{self-consistent}}(t,\b X)-\b F(t,\b X),
\end{equation}

\noindent that we estimate at each iteration through the differential
equation

\begin{equation}
\b w^{(n)}-\b w^{(n-1)}= (a\ddot a+2\dot a^2)
{\mathrm{d}\over \mathrm{d}a}\Delta
\vec F^{(n)}+a\dot a^2{\mathrm{d}^2\over \mathrm{d} a^2}\Delta\vec F^{(n)}.
\end{equation}

\noindent This is however a non-trivial numerical problem that is
beyond the scope of this article and we plan to pursue this possibility
in a future work.
%However, it is interesting to introduce
%an approximation to the iteration procedure. For slowly changing
%velocity fields, we can approximate $\dot{\b u}^{(n)}\approx
%\dot{\b u}^{(1)}$. One expects this approximation to be satisfactory
%for most practical cases, where cosmic fields are smoothed over
%sufficiently large scales. Therefore, we can derive $\b u^{(n)}$ directly from
%$\b w^{(n)}$ without solving (\ref{deltaF}):

%\begin{equation}
%\b u^{(n)}\approx H^{-1}(t)\,(\b w^{(n)}-\dot{\b u}^{(1)}).
%\end{equation}

%\noindent This iterative procedure would, in principle, enable one
%to find a self-consistent over-density field from a current
%peculiar-velocity field
%$\b u^{(1)}$ in the limit $\epsilon \rightarrow 0$.

\section{Comparison with other work}

We have calculated a unique and well-defined relation (\ref{result}) for
cosmic density and velocity fields that is a self-consistent
solution in the second-order model. Given the second-order ansatz for
the particle orbits (\ref{eulerian}), the
Lagrange-Newton system determines the functions of time $f_1(t)$
and $f_2(t)$ for generic initial conditions
and it constrains the perturbation potentials $S^{(1)}$,
$S^{(2)}$ in terms of the peculiar-velocity potential $S$ at $t_0$.
Consequently, the orbits are determined at all times until shell-crossing,
and the density is obtained from the continuity equation by evaluating the
Jacobian $J^{E}$ of the transformation $\b X\rightarrow \b q$.

For simplicity, we have discussed the formulae on the assumption
that the velocity and acceleration fields are parallel.
In Sect. 3, it is shown that this assumption is a sufficient condition
but by no means necessary, and in general it is sufficient to adopt the condition
of parallelism of the gradients of the fields, which is a weaker
requirement.

The self-consistency of the solution for the density field
follows from the fact that the orbits employed are second-order
solutions in perturbation theory and not any given ansatz. Our
result (\ref{result}) generalizes previous results in the literature
in the following respect. Nusser \etal (1991) obtained two
relations for the density constrast in the ZA,

\begin{equation} \label{NusserI}
\delta_c(t,\b q)=-D\;I^E(v_i)+D^2\;II^E(v_i)-D^3\;III^E(v_i),
\end{equation}

\noindent and

\begin{equation} \label{NusserII}
\delta_e(t,\b q)= -D\;I^E(v_i),
\end{equation}

\noindent where a scaled peculiar velocity is used,

\begin{equation}
v_i\equiv (a\dot a)^{-1}u_i, \label{velocity}
\end{equation}

\noindent and $\delta_c$ is derived from the continuity equation and
$\delta_e$ from Euler's equation. The difference between
(\ref{NusserI}) and (\ref{NusserII}) is second order and is
therefore within the order of the approximation:

\begin{equation}
\delta_c(t,\b q)-\delta_e(t,\b q)= D^2\;II^E(v_i)-D^3\;III^E(v_i).
\end{equation}

\noindent Thus, in the ZA, only the first term on the RHS of (\ref{NusserI})
is significant and consistent with conservation of momentum. In the second-order
result (\ref{result}), we are able to retain two more terms and this
relation has a similar form as (\ref{NusserI}), where $D$ is
replaced by $h$. This form of the over-density field satisfies
automatically Euler's equation to second-order, since the orbits
are obtained solving the Lagrange-Newton system to this order, and
therefore the $II(v_i)$ and $III(v_i)$ terms are a legitimate part of
the solution and not higher-order `error' terms as in (\ref{NusserI}).

One does not obtain self-consistent solutions in
models where an arbitrary perturbative parameter, such as $D(t)$,
is singled out, which results in a hybrid Eulerian-Lagrangian scheme.
The chief problem inherent to such approaches is
that the time dependence is set {  ab initio} on the orbits and
it is not determined by the equations of motion as is done
in the solutions we have used.
This type of ansatz can be employed to good effect in a
least-square fit for the
dynamics of the particles as a two-boundary condition problem
(see for example Susperregi \& Binney 1994),
but in perturbation theory they are
inconsistent and of little predictive power due to
their lack of convergence as discussed in the Introduction.
In such approaches, both
the continuity equation and the equation of motion must be truncated
to the given perturbation order and, consequently,
the corresponding solutions $\delta_c$ and $\delta_e$
in general differ. Therefore, from the analysis of the previous section,
we can also conclude that $\delta_c$ and $\delta_e$ do not only differ
by a higher order in the approximation, but failure to solve
for the orbits perturbatively is conducive to discrepancies in
the solutions {\it at the same order}.
Given a parametrization of the orbits such as
(\ref{moutarde}),
the second-order solution for the continuity density is given by
(Gramann 1993b)

\begin{equation}
\delta_c= -D\;I^E(v_i)+D^2\;II^E(v_i),\label{gramann}
\end{equation}

\noindent As shown by Gramann (1993b) the second order term for $\delta_e$
differs by a factor of $4/7$ from that of (\ref{gramann}). This
inconsistency at second order arises as a consequence of
using second-order orbits that do not correspond to the actual
second-order perturbation theory solutions.
In contrast to the solution (\ref{result}), (\ref{gramann})
only accounts for quadratic couplings in $u_i$. Considering that
at first order we already have the relation (\ref{result}), it is
apparent that (\ref{gramann}) is not an accurate generalization of
the Lagrangian linear theory.

In the scheme discussed here, initial conditions are set at $t=t_0$,
and hence we can try to establish a connection with the reconstruction model
(\ref{gramann}) by specifying these at $t=0$. However, the normalization
chosen for $a(t)$ impedes to set $t_0$ identically equal to zero.
A choice of an arbitrarily small $t_0$ requires
keeping only the growing solutions
in the expression for $h(t)$, entirely neglecting the presence of
decaying modes. We have to further restrict ourselves to the leading term
(which is a particular solution of the second-order differential equation)
and neglect the homogeneous second-order solution (which is growing).
The leading order in $h(t)$ is, at both first and second
order (for model I),

\begin{equation}
h\approx {3\over 2}\,t_0\,a^{1/2},
\end{equation}

\noindent and thus, using definition (\ref{velocity}) in (\ref{result}), we get

\[
\begin{array}{rl}
\!\delta\!\!
&= -(ha\dot a)\,I^E(v_i)+(ha\dot a)^2\,II^E(v_i)-(ha\dot a)^3\,III^E(v_i)\\
&\\
&\approx -(a-1)\,I^E(v_i)+(a-1)^2\,II^E(v_i)-(a-1)^3\,III^E(v_i).
\end{array}
\]
\begin{equation}   \label{growing-modes}
\end{equation}

\noindent With the normalization chosen for $a(t)$, (\ref{growing-modes}) is
in reasonable agreement with
(\ref{gramann}) at early times. At late times however,
the presence of the $III^E(v_i)$ term in (\ref{growing-modes}) diminishes the
departures from linearity
brought about by the quadratic term whereas this effect is
not accounted for in (\ref{gramann}).

It must be emphasized that (\ref{result}) is strictly only valid so long as the
inversion $\b H(t,\b q)$ is well-defined. Indeed, the expression given for
$\delta(t, \b H(t,\b q))$ provides the Eulerian values of the density contrast
and is given as a function of Eulerian positions
up to orbit-crossing time. At later times the inversion map becomes singular
and this result does not hold.

By integrating the Poisson equation in Eulerian coordinates
using $\delta(t, \b H(t,\b q)$ it is concluded (e.g. Gramann 1993a,
eq. (2.20)) that the presence of the quadratic term $II^E(v_i)$
in the density-velocity relation induces a departure from
parallelism between peculiar-velocity and -accel\-eration vectors.
The derivation of this relation under the assumption of parallelism
proves this conclusion is not correct. Certainly, one would
be able to obtain agreement with solutions of Poisson's
equation if a self-consistent density contrast were known. It is
manifest however that this is far from being the case in these
models. Differences in the solutions within the order of the
approximation, as it is the case with $\delta_c$ and $\delta_e$
in Gramann's analysis, imply that any conclusion drawn from Poisson's
equation applied to the reconstructed density
is not justified as long as approximations are involved.

We finally note that an expansion of the Jacobian to a given order
({\it after} the solutions are obtained) mirrors formally the
mathematical fact that, in principle, we cannot believe terms
of higher order than the perturbative order of the trajectories.
However, the powerful properties of a Lagrangian approach
mainly rely on the existence of an exact integral for the density
for {\it any} given trajectory field, which allows using these approximations
until shell-crossing. At this epoch the contributions from second
and third invariants is of the same order as that from
the first invariant, a fact which underlies the success of
the ZA.
Whether this extrapolation until shell-crossing is justified at all
is another issue. The successes of Lagrangian perturbation methods
as summarized by Bouchet et al. (1995) and the success of
following trajectories until shell-crossing and even beyond
(Melott et al. 1995) strongly argue in favour of this extrapolation.

\section{Conclusions}

In this paper we have presented a relation between cosmic density and
velocity fields following a purely Lagrangian derivation (Eq.(\ref{result}))
that is self-consistent in the second-order model. The advantage
of working in the Lagrangian picture is that the over-density field
obtained from the continuity equation is self-consistent and automatically
satisfies conservation of momentum.
For simplicity, one can adopt the
condition (\ref{parallelism}) of parallelism of the fields $\b u$
and $\b w$, satisfied at first order
and by a large class of irrotational solutions in second-order Lagrangian
perturbation theory, and the result (\ref{result}) holds accurately at both orders.
However, as pointed out in Sect. 3, we note that (\ref{parallelism}) is
not a sine qua non condition but a weaker condition of parallelism of the
gradients of the fields would yield the same result. It is shown that
this condition is well satisfied for generic initial conditions.
In each case, the function $h(t)$ is given explicitly in
Appendix B.1 and Appendix B.2 respectively.
Its range of validity is limited by the epoch when
the inversion map $\b H(t,\b q)$ becomes singular, i.e. up to orbit-crossing
time, while its principal range of validity should strictly be estimated by
$|p_{i,j} (t_c)|<< a(t_c)$,
where $\b p(t,\b X) = a(t)\,[\b F(t,\b X) - \b X]$ is
the displacement field with respect to the Hubble-flow.
This condition is probably very conservative in view of the
success of these approximations if followed up to shell-crossing.

By specifying the initial data at $t\approx 0$ and keeping only the leading
terms in the solutions, we are able to recover the leading terms of
previous reconstruction models as shown in Sect. 5. Furthermore, the
time dependence of the orbits in our model is established by solving
the Lagrange-Newton system at each order.
This leads to a unique and consistent
result for the Eulerian density constrast in terms of the
peculiar-velocities.

We finally wish to show some skepticism about the usefulness
of the reconstruction formulae given. We want them to be
applicable to present day observational data. However, we
know confidently that shell-crossing has occurred, i.e.
the interesting regime is no longer covered.
Instead one would have to investigate a relationship between
{\it smoothed} cosmic over-density and peculiar-velocity fields, where the
smoothing window scale has to ensure the absence of vorticity and
multi-stream systems in the average flow.
Given the present work we should ask whether there is a
formally proper way to incorporate smoothing into a consistent
Lagrangian reconstruction formalism. With regard to this point,
we think that solutions of the Lagrange-Newton system
may possibly provide reasonable tools for reconstructing average fields, since
averages in Eulerian space of any vector function
$\b A (t,\b H (t, \b q))$ are straightforward to calculate:
$\langle\b A\rangle_{q} \propto \int d^3 \b q \; \b A (t,\b X) =
\int d^3 \b X \; J (t,\b X) \; \b A (t,\b X)$.
However, we still
encounter the fundamental problem of averaging over multi-stream flows,
which may not be properly handled in the framework of the
Lagrange-Newton system. Rather, such a description has to be
based on an approximation of the Vlasov-Poisson system.
This major improvement of the methods lies well beyond the scope
of present  reconstruction techniques and requires the construction
of Vlasov-Poisson type approximations.

\begin{acknowledgements}
We thank James Binney, Avishai Dekel,
J\"urgen Ehlers, Adi Nusser and David Weinberg
for valuable comments and suggestions. TB is supported by the
"Sonderforschungsbereich 375-95 f\"ur
Astro-Teilchenphysik" der Deut\-schen Forschungsgemeinschaft.
MPS is grateful to the Max-Planck-Institut
f\"ur Astrophysik at Garching for its hospitality.
\end{acknowledgements}

\appendix
\section{determination of $h(t)$}

\subsection{First-order model I}

The first-order particle orbits are

\begin{equation}
\b F(t,\b X)= \b X+f_1(t)\,t_0\,\nabla_X S (\b X),  \label{firstorder}
\end{equation}

\noindent where the perturbation potential is related to
the initial gravitational potential through the relation

\begin{equation}
\nabla_{\b X} S = -t_0^2\;\nabla_{\b X}\phi.
\end{equation}

\noindent The inversion of (\ref{firstorder}) is then:

\begin{equation}
\b H(t,\b q)= \b q -h(t)\;\b u(t,\b H)
=\b q -a^{-1/2}\,t_0\,f_1(t)\;\b u(t,\b H),
\end{equation}

\noindent and therefore $h(t)$ is given by:

\begin{equation}
h={3\over 2}\,t_0\, (a^{1/2}-a^{-1/2}). \label{1h}
\end{equation}

Thus, we obtain:

\begin{equation}
\b u= a^{1/2}\,t_0^{-1}\;\nabla_{\b X} S,
\end{equation}

\begin{equation}
\b w\, t_0= a^{-1}\,t_0^{-1}\; \nabla_{\b X} S.
\end{equation}

\noindent From this we derive the well-known result for the Zel'dovich
approximation used in the Sachs-Wolfe effect that

\begin{equation}
\nabla_q \phi ={\mathrm{const}},
\end{equation}
since

\begin{equation}
\b w= -{1\over a\,t_0^2} \nabla_q \phi.
\end{equation}

\subsection{Second-order model I}

\noindent Eqs. (\ref{u})(\ref{w}) in terms of the perturbation potentials
read

\begin{equation}  \label{uwmatrix}
\left[\begin{array}{c}
 \b u(t,\b H) \\ \b w(t,\b H)
 \end{array}\right]= \left[\begin{array}{cc}
     \alpha(t) & \beta(t) \\
      \eta(t) & \zeta(t)
     \end{array}\right] \left[\begin{array}{c}
                             \nabla_X S^{(1)}(\b X)\\
                              \nabla_X S^{(2)}(\b X)
                         \end{array}\right],
\end{equation}

\noindent where

\begin{equation}
\alpha= t_0^{-1}\,a^{1/2},
\end{equation}

\begin{equation}
\beta= {3\over 2}\,t_0^{-1}\;\Big(-{3\over 7}a^{3/2}+{3\over 5}a^{1/2}
-{6\over 35}a^{-2}\Big),
\end{equation}

\begin{equation}
\eta=  t_0^{-2}\;a^{-1},
\end{equation}

\begin{equation}
\zeta= t_0^{-2}\Big(-{15\over 14}+{9\over 10}a^{-1}+{6\over 35}a^{-7/2}\Big).
\end{equation}

\noindent We write the perturbation potentials in terms of $\b u$ and $\b w$ by
inverting (\ref{uwmatrix}):

\begin{equation} \label{smatrix}
 \left[\begin{array}{c}
 \nabla_{X} S^{(1)}(\b X) \\
 \nabla_{X} S^{(2)}(\b X)
 \end{array}\right]= \left[\begin{array}{cc}
     \alpha'(t) & \beta'(t) \\
      \eta'(t) & \zeta'(t)
     \end{array}\right] \left[\begin{array}{c}
                             \b u(t,\b H)\\
                             \b w(t,\b H)
                         \end{array}\right],
\end{equation}

\noindent where

\begin{equation}
 \alpha'=(1-a^{-7/2})^{-1}t_0\,({5\over 2} a^{-1/2}-{21\over 10}a^{-3/2}
-{2\over 5}a^{-4}),
\end{equation}

\begin{equation}
 \beta'=(1-a^{-7/2})^{-1}t_0^2\,(-{3\over 2}a+{21\over 10}-{3\over 5}a^{-5/2}),
\end{equation}

\begin{equation}
 \eta'= {7\over 3}(1-a^{-7/2})^{-1}t_0\,a^{-3/2},
\end{equation}

\begin{equation}
\zeta'= -{7\over 3}(1-a^{-7/2})^{-1}t_0^2.
\end{equation}

\noindent Using (\ref{smatrix}), we invert (\ref{eulerian}) in the form

\begin{equation}
\b H(t,\b q)=\b q-(1-a^{-7/2})^{-1}[\,m_1(t)\,\b u +
m_2(t)\, \b w \,], \label{inverted}
\end{equation}

\noindent where

\begin{equation}
m_1= {3\over 2}\,t_0\;( {7\over 4}a^{1/2}-{5\over 2}a^{-1/2}
+{7\over 20}a^{-3/2}+{2\over 5}a^{-4}),
\end{equation}

\begin{equation}
m_2={3\over 2}\,t_0^2\;(-{3\over 4}a^2
+{3\over 2}a-{7\over 20}-a^{-3/2}+{3\over 5}a^{-5/2}).
\end{equation}

\noindent Therefore, given the condition of parallelism(\ref{parallelism}),
we have that (\ref{inverted}) reduces to (\ref{lagrangian2}) with

\begin{equation}
h(t)\equiv (1-a^{-7/2})^{-1}[\,m_1(t)+t^{-1}\,m_2(t)\,]
 \end{equation}

\begin{equation}
h= {3\over 2}\,t_0\;(1-a^{-7/2})^{-1}(a^{1/2}-a^{-1/2}-a^{-3}+a^{-4}).
\label{hsecond}
\end{equation}

\subsection{Second-order model II}

Following the same procedure as in Appendix A.2, we write

\begin{equation}  \label{uwmatrix2}
\left[\begin{array}{c}
 \b u(t,\b H) \\ \b w(t,\b H)
 \end{array}\right]= \left[\begin{array}{cc}
     \alpha(t) & \beta(t) \\
      \eta(t) & \zeta(t)
     \end{array}\right] \left[\begin{array}{c}
                             \nabla_X S^{(1)}(\b X)\\
                              \nabla_X S^{(2)}(\b X)
                         \end{array}\right],
\end{equation}

\noindent where

\begin{equation}
\alpha={2\over 5}\,t_0^{-1}\, (a^{1/2}+{3\over 2}a^{-2}),
\end{equation}

\begin{equation}
\beta= {9\over 25}\,t_0^{-1}\,(-{2\over 7}a^{3/2}+{29\over 60} a^{1/2}+
{1\over 6}a^{-1}-{43\over 70}a^{-2}+{1\over 4}a^{-7/2}),
\end{equation}

\begin{equation}
\eta=  {2\over 5}\,t_0^{-2}\,(a^{-1}-a^{-7/2}),
\end{equation}

\begin{equation}
\zeta= {3\over 25}\,t_0^{-2}\,(-{10\over 7}+{29\over 20}a^{-1}+{43\over 35}a^{-7/2}
-{5\over 4}a^{-5}).
\end{equation}

\noindent Therefore, inversion of (\ref{uwmatrix2}) leads to:

\begin{equation}
 \left[\begin{array}{c}
 \nabla_{X} S^{(1)}(\b X) \\
 \nabla_{X} S^{(2)}(\b X)
 \end{array}\right]=\gamma (t)\, \left[\begin{array}{cc}
     \alpha'(t) & \beta'(t) \\
      \eta'(t) & \zeta'(t)
     \end{array}\right] \left[\begin{array}{c}
                             \b u(t,\b H)\\
                             \b w(t,\b H)
                         \end{array}\right],
\end{equation}

\noindent where

\begin{equation}
 \alpha'=28\,t_0\, a^2,
\end{equation}
\[
 \beta'=
\]
\begin{equation}
\hspace{0.5cm} -{9\over 5}\,t_0^2\,\Big(-{2\over 7}a^{3/2}
+{29\over 60}a^{1/2}+{1\over 6}
a^{-1}-{43\over 70}a^{-2}+{1\over 4}a^{-7/2}\Big),
\end{equation}

\begin{equation}
 \eta'= 2\,t_0\,(-a^{-1}+a^{-7/2}),
\end{equation}

\begin{equation}
\zeta' = t_0^2\,(2a^{1/2}+3a^{-2}),
\end{equation}

\[
  \gamma =
\]
\begin{equation}
\hspace{0.2cm}  \Big(-{24\over 175}a^{1/2}-{21\over 25}a^{-2}+{45\over
28}a^{-3}
-{9\over 25} a^{-9/2}-{27\over 100}a^{-7}\Big)^{-1}.
\end{equation}

\section{Jacobian $J^E$}
\renewcommand{\theequation}
{B\arabic{equation}}
\setcounter{equation}{0}

The Jacobian $\det(\partial H_i/\partial q_j)$ is given by:
\[
  \begin{array}{ll}
J^E =& \!\!1 - g_1\,I^E(u_i) - g_2\,I^E(w_i) +g_1^2 \,II^E(u_i)
+ g_2^2\, II(w_i)\\

  &\!\!+g_1\;g_2\; II'^E(u_i\mid w_j) - g_1^3\; III^E(u_i) - g_2^3\; III^E(w_i)\\

  &\!\!-g_1^2\; g_2\; III'^E(u_i\mid w_j)-g_1\;g_2^2 \;III''^E(u_i\mid w_j),
\end{array}
\]
\begin{equation}   \label{full-jacobian}
\end{equation}

\noindent where

\begin{equation}
I^E(u_i)=u_{i,i},
\end{equation}

\begin{equation}
II^E(u_i)= {1\over 2}(u_{i,i} u_{j,j}- u_{i,j} u_{j,i}),
\end{equation}

\begin{equation}
II'^E(u_i\mid w_j)= {1\over 2}(u_{i,i}w_{j,j}-u_{i,j}w_{j,i}),
\end{equation}

\begin{equation}
III^E(u_i)=\det(u_{i,j})
\end{equation}

\[
\begin{array}{ll}
III'^E(u_i\mid w_j)=\\
u_{2,3} u_{3,2} w_{1,1} - u_{2,2} u_{3,3} w_{1,1} - u_{2,3} u_{3,1}w_{1,2}
+ u_{2,1} u_{3,3} w_{1,2} \\
+ u_{2,2} u_{3,1} w_{1,3} - u_{2,1} u_{3,2} w_{1,3}
- u_{1,3} u_{3,2} w_{2,1} + u_{1,2} u_{3,3} w_{2,1}\\
+ u_{1,3} u_{3,1} w_{2,2}- u_{1,1} u_{3,3} w_{2,2} - u_{1,2} u_{3,1} w_{2,3}
+ u_{1,1} u_{3,2} w_{2,3},
\end{array}
\]
\begin{equation}
\end{equation}

\[
\begin{array}{@{}ll}
III''^E(u_i\mid w_j)=\\
u_{3,3} w_{1,2} w_{2,1} - u_{3,2} w_{1,3} w_{2,1} - u_{3,3} w_{1,1}w_{2,2}
+ u_{3,1} w_{1,3} w_{2,2}\\
+ u_{3,2} w_{1,1} w_{2,3} - u_{3,1} w_{1,2} w_{2,3}
- u_{2,3} w_{1,2} w_{3,1} + u_{2,2} w_{1,3} w_{3,1}\\
+ u_{1,3} w_{2,2} w_{3,1} - u_{1,2} w_{2,3} w_{3,1}
+ u_{2,3} w_{1,1} w_{3,2} - u_{2,1} w_{1,3} w_{3,2},
\end{array}
\]
\begin{equation}
\end{equation}

\noindent and the functions $g_1(t)$ and $g_2(t)$ are model-dependent.
For instance, for the case of model I of Sect. 3,

\begin{equation}
g_1(t)=(1-a^{-7/2})^{-1}\;m_1(t),
\end{equation}

\begin{equation}
g_2(t)=(1-a^{-7/2})^{-1}\;m_2(t),
\end{equation}

\noindent where $m_1(t)$ and $m_2(t)$ are defined in Appendix A.2.
The assumption of parallelism (\ref{parallelism}) adopted,
$J^E$ is simplified by making the following substitutions in (\ref{full-jacobian}):

\begin{equation}
 \begin{array}{lcl}
    \b w &\rightarrow & 0\\
     g_1(t)&\rightarrow &h(t),
  \end{array}
\end{equation}

\noindent where

\begin{equation}
h(t)=g_1(t)+t^{-1}\,g_2(t).
\end{equation}

Hence, all terms containing $w_i$ are set equal to zero,
and thus (\ref{full-jacobian}) simply reads

\begin{equation}
  J^E=1-h\;I^E(u_i)+h^2\;II^E(u_i)-h^3\;III^E(u_i).
\end{equation}


\begin{thebibliography}{}

\bibitem{ref1}
Bagla J.S., Padmanabhan T., 1996, ApJ 469, 480

\bibitem{ref2}
 Bernardeau F., 1992, ApJ 390, L61

\bibitem{ref3}
 Bertschinger  E., Dekel A., 1989, ApJ 336,
L5

\bibitem{ref4}
 Bertschinger E., Dekel A., Faber S.M., Dressler A., Burnstein D.,
1990, ApJ 364 370

\bibitem{ref5}
 Bouchet F.R., Juszkiewicz R., Colombi S., Pellat R., 1992, ApJ 394, L5

\bibitem{ref6}
 Bouchet F.R., Colombi S., Hivon E., Juszkiewicz R.,
 1995, A\&A, 296, 575

\bibitem{ref7}
 Buchert T., 1989, A\&A 223, 9

\bibitem{ref8}
 Buchert T., 1992, MNRAS 254, 729

\bibitem{ref9}
 Buchert T., 1993, A\&A 267, L51

\bibitem{ref10}
 Buchert T., 1994, MNRAS 267, 811

\bibitem{ref11}
 Buchert T., Ehlers J., 1993, (BE93) MNRAS 264, 375

\bibitem{ref12}
 Buchert T., G\"otz, 1987, J. Math. Phys. 28, 11

\bibitem{ref13}
 Catelan P., 1995, {  MNRAS}, 276, 115

\bibitem{ref14}
 Dekel A., 1994, ARA\&A 32, 371-418

\bibitem{ref15}
 Dekel A., Bertschinger E., Faber S.M., 1990, ApJ
{  364}, 349

\bibitem{ref16}
 Doroshkevich A.G., Ryaben'ky V.S., Shandarin S.F., 1973,
Astrophysics 39, 144

\bibitem{ref17}
 Ehlers J., Buchert T., 1997, GRG, in press

\bibitem{ref18}
 Gramann M., 1993a, ApJ 405, 449

\bibitem{ref19}
 Gramann M., 1993b, ApJ 405, L47

\bibitem{ref20}
 Melott A.L., Buchert T., Wei{\ss} A.G., 1995,
A\&A 294, 345

\bibitem{ref21}
 Moutarde F., Alimi J.M., Bouchet F.R., Pellat R., Ramani A., 1991,
ApJ 382, 377

\bibitem{ref22}
 Nusser A., Dekel A., Bertschinger E., Blumenthal G. R., 1991,
ApJ 379, 6

\bibitem{ref23}
 Nusser A., Dekel A., 1992, ApJ  391, 443

\bibitem{ref24}
 Susperregi M., Binney J., 1994, MNRAS 271, 719



\end{thebibliography}
\end{document}